\documentclass[aps,twocolumn,nofootinbib,
preprintnumbers]{revtex4-1}
\usepackage{bm}% bold math
\usepackage{hyperref}% Auto-generate hyperlinks
\usepackage{amsmath,amssymb,amsfonts,graphicx,float}
\renewcommand{\vec}[1]{{\mathbf{#1}}}
\newcommand{\beq}{\begin{eqnarray}}
\newcommand{\eeq}{\end{eqnarray}}
\usepackage{amsmath}
\usepackage{amssymb}
\usepackage[paper=letterpaper,margin=1in]{geometry}
\usepackage{braket}
\usepackage{upgreek}

\renewcommand{\vec}[1]{\boldsymbol{#1}}

\def\a{\alpha}

\def\del{\partial}

\usepackage{tikz}
\begin{document}

\title{Un-Fermi Liquids: Unparticles in Strongly Correlated Electron Matter}
\author{ Philip W. Phillips and Brandon W. Langley}

\affiliation{Department of Physics and Institute for Condensed Matter Theory,
University of Illinois
1110 W. Green Street, Urbana, IL 61801, U.S.A.}

\author{Jimmy A. Hutasoit}

\affiliation{Department of Physics,
The Pennsylvania State University,
University Park, PA 16802 }

\date{\today}

\begin{abstract}

Since any non-trivial infrared dynamics in strongly
correlated electron matter must be controlled by a critical fixed point, we
argue that the form of the single-particle propagator can be deduced
simply by imposing scale invariance.  As a consequence, the unparticle
picture proposed by Georgi\cite{georgi} is the natural candidate to describe such dynamics.
Unparticle stuff is scale-invariant matter with no particular
mass. Scale invariance dictates that the propagator has an algebraic
form which can admit zeros and hence is a candidate to explain the
ubiquitous pseudogap state of the cuprates.   The non-perturbative electronic state
formed out of unparticles we refer to as an un-Fermi liquid.  We show
that the underlying action of the continuous mass formulation of
unparticles can be recast as an action in anti de Sitter
space which serves as the generating functional for the propagator.  We find that this mapping fixes the scaling dimension of
the unparticle to be $d_U=d/2+\sqrt{d^2+4}/2$ and ensures that the
corresponding propagator has zeros with $d$ the spacetime dimension of
the unparticle field. Should $d=2+1$, unparticles acquire the
non-trivial phase $2\pi d_U$ upon interchange.  Because $d_U$ is
non-integer and in general not half-integer, clockwise and counterclockwise interchange of unparticles
do not lead to the same phase and time reversal symmetry is broken
spontaneously as reported in numerous experiments in the pseudogap
phase of the cuprates.  The possible relevance of this mechanism to
such experiments is discussed.  We then formulate the
analogous BCS gap using unparticles and find that in contrast to the
Fermi liquid case, the transition temperature increases as the
attractive interaction strength decreases, indicating that unparticles
are highly susceptible to a superconducting instability.  
\end{abstract}

\maketitle

\section{Introduction}

The key problem that arises from the strong correlations in the normal
state of the copper-oxide superconductors is  identifying the weakly
interacting entities
that make a particle interpretation of the current possible.
Superconductivity would then be reduced to a pairing
instability of such objects.  However, there is good reason to believe
that the construction of such entities may not be possible.  The
reason lies in the fact that both the parent and pseudogap phases are
characterized by a vanishing of the single-particle propagator
$G(E,\vec p)$,
evaluated at zero energy\cite{johnson,kotliar,imada,tsvelik,dzy} .   While ${\rm Det}[G(E=0,\vec p)]=0$ \footnote{This condition is not satisfied in phenomenological accounts based on mean-field theory.} for certain momenta in
the pseudogap state, the single-particle propagator vanishes for all
momenta in the parent Mott
insulating state\cite{dzy,tsvelik,phillips} as would be expected for a hard gap in the absence of
symmetry breaking.  If we write the propagator simply as
$1/(E-\varepsilon_{\vec p}-\Sigma(E,\vec p))$, a vanishing propagator
(where both the real and imaginary parts vanish) is
possible only if the self-energy diverges.  Because the imaginary part
of the self-energy defines the decay rate of a state, a divergent self
energy implies that no stable particle-like excitation is present. In
fact, to date the only known kind of excitations that emerge from
zeros of a single-particle propagator are bound or composite states,
for example Cooper pairs in the context of superconductivity, which do not admit a particle interpretation in terms of a
quadratic action\footnote{While it would be convenient for a quadratic
  action to underlie zeros, such an occurrence would imply a particle
  interpretation of zeros, which is a direct contradiction.} with canonical fields.
Nonetheless, a highly influential result, the Luttinger
count\cite{luttinger}, which has been applied widely in the field of strong correlations\cite{yrz,fahrid,tsvelik},
equates the number of excitations which can be given a particle interpretation
not only with the number of poles, or quasiparticles, but also with the
number of zeros of the
single-particle Green function.   If this were true, this would be
truly remarkable as it would imply that even in the limit
diametrically opposed to the Fermi liquid or quasiparticle regime,
the particle concept still applies.  Hence, on physical grounds, equating
the Luttinger count with the particle density is difficult to fathom and inconsistent with the
work of 't Hooft\cite{thooft} who has shown that the analogous problem in
QCD in $d=1+1$  implies that there are ``no physical
quark states'' at low energies. 

Since we believe that strongly correlated electron matter is
no different, one of us\cite{dave} analyzed the precise
mathematical statement underlying Luttinger's claim,
\beq\label{nsumg}
n_L=2\sum_{\vec p}\Theta({\rm Re}G(\omega=0,\vec p)),
\eeq
with
$G(\omega,\vec p)$ the single-particle propagator, $\Theta$ the Heaviside step function and
$\vec p$ the momentum, and has shown that it is, understandably, false in the sense
that when zeros are present, $n_L$ is not necessarily the particle
density.  The failure of Eq. (\ref{nsumg}) to yield the particle
density in the normal state of the cuprates where zeros are present has a profound consequence.
The right-hand side of Eq. (\ref{nsumg}) is evaluated strictly at
zero-energy or at the chemical potential.  Hence, if Eq. (\ref{nsumg})
fails to yield the charge density, there are charged degrees of
freedom left over which have no interpretation in terms of low-energy physics alone,
that is particles.  

Consequently, the key question that arises is: What is
the stuff in the normal state of the cuprates (or more generally
strongly correlated electron matter) that couples to the current but has no
particle interpretation?  
We put forth here that the unparticle stuff proposed by H. Georgi\cite{georgi}
several years ago provides a reasonable answer to this problem.  We
stick with the characterization used by Georgi\cite{georgi} that
unparticles should be called stuff because all other characterizations
imply, mistakenly, a particle correspondence.  There is strictly none to be had for
unparticle matter.  Unparticle stuff is scale invariant matter with no
particular mass but with non-trivial (non-Gaussian in terms of
canonical fields) IR dynamics. This construction is
natural in strongly correlated systems because the interactions exist
on all length scales.  A key prediction of this correspondence is that
strongly correlated electron matter that has a vanishing propagator
exhibits fractional statistics in $d=2+1$, which is experimentally falsifiable. 

\section{ Precursors}

We motivate the relevance of unparticle stuff to strongly correlated
electron matter by recalling what happens when the Wilsonian procedure
is carried out on the basic model for a doped Mott insulator, namely, the Hubbard model.  In this model, electrons hop among neighboring lattice sites with
amplitude $-t$ and encounter an on-site repulsion of magnitude $U$
when two electrons with opposite spin occupy the same site. As $U$ is
the high-energy scale, the goal of a Wilsonian procedure is to
integrate out the $U$-scale physics.  However, this is not simple to
do because the physics at the $U$-scale involves a 4-fermion term.
However, we have shown how this can be done exactly. The procedure is
well documented\cite{ftm,ftm2,ftm3} so we will just recount the
essential parts.  

The key idea\cite{ftm,ftm2,ftm3} is to extend the
Hilbert space by introducing a new fermionic operator, $D_i^\dagger$,
which permits a clean identification of the $U$-scale physics.  The operator $D_i$ 
enters the $UV$- complete Lagrangian with a mass of $U$.  However, it
only corresponds to the creation of double occupancy through a
constraint, 
$\delta(D_i-\theta c_{i\uparrow}c_{i\downarrow})$, where $\theta$ is a
Grassmann variable and $c_{i\sigma}$ is the electron annihilation operator for
site $i$ with spin $\sigma$.  In essence, we have fermionized double
occupancy by the introduction of the Grassmann field and hence, $D_i$
to some extent represents a super field.  The constraint is imposed with a Lagrange multiplier,
$\varphi_i$, which must have charge $2e$ because $D_i$ has charge $2e$. The $UV$-complete theory written in terms of the
$D_i$ field is formally equivalent to the Hubbard model: integration
over $\varphi_i$ and then integration over $D_i$ results in the action
for the Hubbard model.  However,
equivalence at the UV scale is not the key point here.  The reason for
adopting this new language is to be able to integrate out the $U$-scale physics exactly.  Because $D_i$ is a canonical fermionic field
and it enters the action in a quadratic fashion,
it can be integrated out exactly. The Lagrangian that results will
describe the IR physics and will depend on the constraint field
$\varphi_i$. 

We note that in the integration
over the $U$-scale physics, $D_i$ represents whatever the physics is
on the $U$-scale.  Only in the atomic limit is this strictly double
occupancy.  Similarly, in the IR theory, the emergent field $\varphi_i$
also will not represent double occupancy but whatever physics the
upper Hubbard band produces in the lower band by virtue of dynamical
spectral weight transfer.  When all of this is done, what is most
important here is that  one can
identify what the charge degrees of freedom look like in the IR by
adding a minimally coupled source term to the $UV$-complete theory.
The source term acts in the extended space and is carefully chosen so
that when the constraint is solved, the bare electron operator is then
minimally coupled to the source term.  

However, integrating out the $D_i$
field tells another story.  The new IR charge which is now minimally
coupled to the source, let us call it $\psi_{i\sigma}$, 
depends explicitly on $\varphi_i$.  The square of this field, which
defines the number density of such excitations, can be
computed explicitly\cite{ftm4}.  For a lower Hubbard band with $x$ holes, the
conserved charge is $1-x$.   Nonetheless, $\langle |\psi_{i\sigma}|^2\rangle<1-x$.
The deficit corresponds to all the stuff that couples to $\varphi_i$,
the Lagrange field that tethers $D_i$ to the UV scale.  Hence, all
of the charge degrees of freedom which depend on $\varphi_i$ contribute
to the current but \emph{not} to the particle density.  In fact, they create zeros of
the single-particle Green function\cite{ftm4} and hence cannot be
given a particle interpretation, thus their vanishing contribution to
the particle density.  

This implies that the Wilsonian procedure on the Hubbard model
provides a clear example of an emergent IR theory with charged degrees
of freedom that have no particle (electron)  interpretation.  In fact,
the culprit emerges from the effective interpolating\cite{jimmy} field, $\varphi_i$,
which is made manifest in the IR theory entirely by eliminating the
$UV$-scale physics.  It couples to particle stuff and leads to zeros of the single-particle Green function\cite{ftm4}.  In the context of physics beyond
the standard model, Georgi\cite{georgi} has proposed that such fields
can arise and lead to non-trivial IR dynamics by interacting with
the particle sector, precisely in the manner found here.  Such fields
generate\cite{jimmy} unparticle physics. 

That such physics should enter the Hubbard model can be seen by
comparing the number of electron states at low energy with the total
spectral weight. In a Fermi liquid, the two are necessarily equal, but it is not
so in an expansion around the atomic limit of the Hubbard model.  The
number of electron states is delineated from a pure stoichiometric
argument, namely counting the number of sites.  The Mott state
corresponds to $1$-electron per site.  Hence, there are $N$ electron
addition and $N$ electron removal states in an $N$-site system.  Per
site, this simply means that there is a single removal and addition
state.  Let $x=n_h/N$, where $x$ is the number of empty sites.  The
number of electron removal states when $x$ holes are present is
$1-x$.  Each
hole can be filled with either a spin-up or a spin-down electron.
Hence, $x$ holes produce $2x$ electron states, which lie at low energy.  Therefore, the number of low-energy
electron states is just a sum of the electron removal plus the number
of hole states,  $1-x+2x=1+x$. This number is not affected by the
dynamics.  However, the spectral weight does change as this is
determined by the true propagating degrees of freedom in the IR and
hence the dynamics.   As shown many
years ago\cite{harris} and observed
experimentally\cite{ctchen,sawatzky}, the spectral weight in the lower
band is given by $1+x+f(x,t/U)$
where $f(x,t/U)>0$.   Consequently, counting electrons cannot exhaust
the number of degrees of freedom in the lower band.  The missing
degrees of freedom do not have a particle interpretation.  Furthermore, such degrees of freedom play a crucial role
in the doped system as they provide a mechanism for zeros of the
single-particle propagator, the key mechanism for the breakdown of
Luttinger's theorem. Further, it has been suggested\cite{sawatzky1} (without proof in the last sentence of the paper) that dynamical spectral weight transfer can only
be captured by a low-energy theory if the fundamental excitations  in the IR have fractional statistics.  As will be seen, this feature appears naturally in our low-energy construction.

\section{Unparticles}

\begin{figure}
\begin{center}
\includegraphics[width=3.0in]{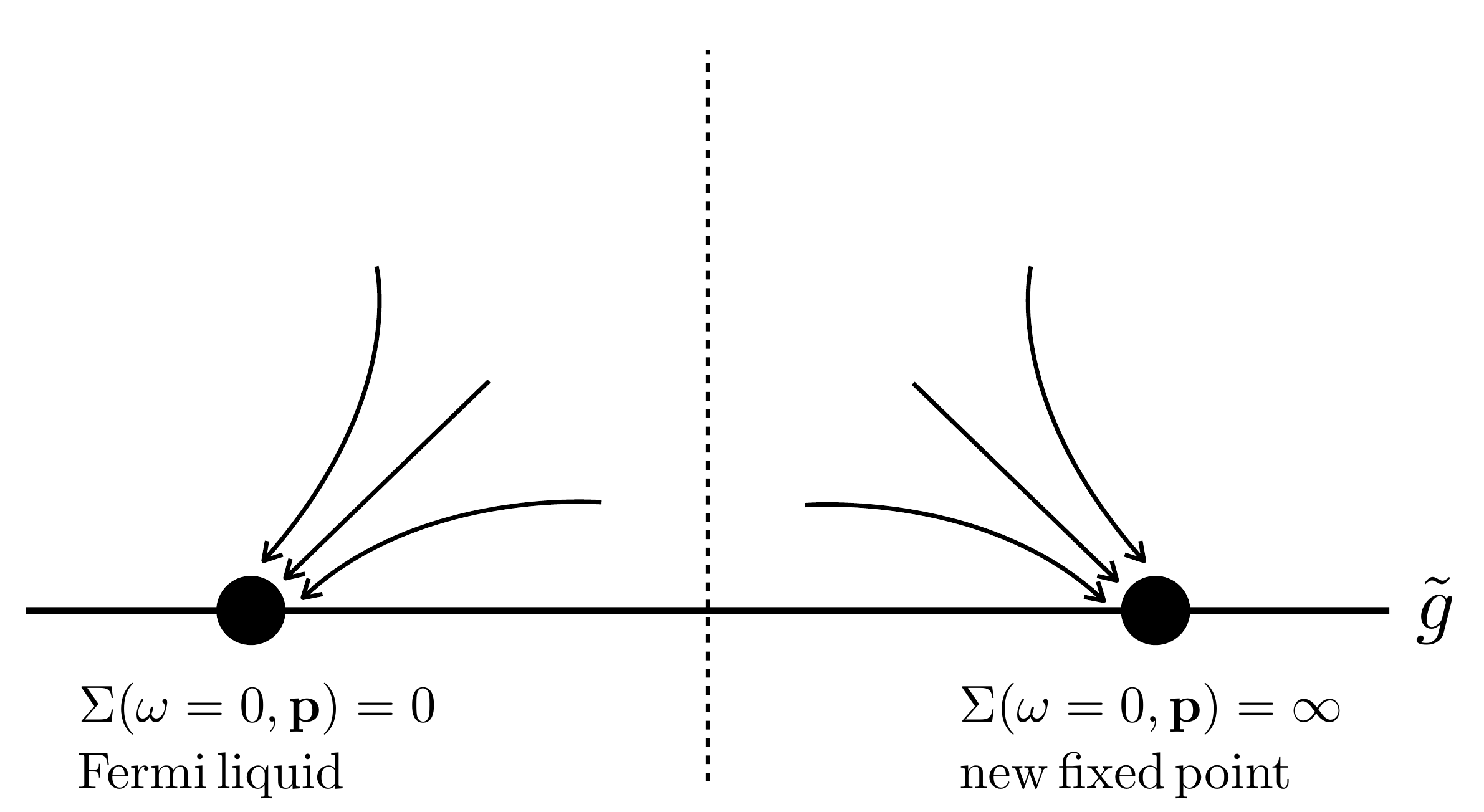}
\caption{  Heuristic renormalization group flow for the Fermi liquid fixed point in which the self-energy is zero or negligible in the IR and
one in which zeros of the single-particle propagator appear.  Since the self energy diverges in the latter, the excitations which appear here are not adiabatically connected to the Fermi liquid fixed point.  The breakdown of the particle concept at the new fixed point suggests an unparticle picture is valid.  The horizontal axis represents the strength of the coupling constant.}
\label{rgpict}
\end{center}
\end{figure}

The breakdown of the Fermi-liquid picture in a doped Mott insulator
stems generically from a divergent self energy at zero energy (see Fig. \ref{rgpict}).  Any
excitations that arise from such a divergence are clearly not
adiabatically connected to the non-interacting or Fermi-liquid fixed point.
Consequently, if any new excitations emerge, they must arise
fundamentally from a new fixed point as illustrated in Fig. \ref{rgpict}. While it is difficult to establish the existence of a non-trivial
(non-Gaussian in the UV variables) IR fixed point of the Hubbard model (or any strongly
coupled model for that matter), the cuprates, Mott systems in general which display quantum critical scaling\cite{kanoda,mc1} and the Wilsonian
procedure for the Hubbard model
suggest that such a non-trivial fixed point emerges out of the strong
relevant on-site interactions, evidenced in part by the numerous
experimental signatures\cite{marel,valla,mfl} of scale invariance in the normal
state as well as highly successful phenomenology\cite{mfl} that relies on such
invariance.  What permits immediate quantitative progress here is that all critical
fixed points exhibit scale invariance and this principle
anchors fundamentally the kind of single-particle propagators that arise as pointed
out by Georgi\cite{georgi}.  Since mass and scale invariance are incompatible, the excitations
which emerge have no particular mass and are called unparticles\cite{georgi}. The propagator\cite{georgi} in the
scalar unparticle sector
can be written down strictly from scale invariance
\beq\label{gfunparticles}
G_U(\vec{k})&=& \frac{A_{d_U}}{2\sin(d_U\pi)}\frac{i}{(
  \vec{k}^2-i\epsilon)^{d/2-d_U}},\nonumber\\
 A_{d_U}&=&\frac{16\pi^{5/2}}{(2\pi)^{2d_U}}\frac{\Gamma(d_U+1/2)}{\Gamma(d_U-1)\Gamma(2d_U)},
\eeq
where $ \vec k$ is the $d$-momentum, $d$ is the dimension of the spacetime the unparticle lives in and $d_U$ is the scaling dimension
of the unparticle operator, typically not an integer. Here, we adopt the notation where the diagonal entries of the metric are mostly positive. 

Consequently, the failure of
the particle concept is manifest by the branch cut in $G_U$.  As a
result, $G_U$ bears some resemblance to the propagator for Luttinger
liquids\cite{volovik}.  However, $G_U$ lacks the topological term in
the denominator that preserves the singularity at
the Fermi momentum, $k_F$, ultimately the key to the successful
implementation of the bosonization
scheme for  Luttinger liquids.

Our precise claim is that the low-energy physics of strongly
correlated electron systems, at least the sector governed by the
interpolating field, $\varphi_i$ in the context of the Hubbard model,
is described by a spectral function that scales with an anomalous
exponent. Since $\varphi_i$ is only present when the hopping matrix
element is non-zero,
our analysis does not apply to the atomic limit in which the Green
function can be obtained exactly.  Within the unparticle
proposal, the spectral function should have the form
\beq\label{specfun}
A(\Lambda\omega,\Lambda^{\a_{ \vec{k}}} \vec{k})&=&\Lambda^{\alpha_A}A(\omega,\vec k),\nonumber\\
A(\omega,\vec{k}) &=& \omega^{\alpha_A}f_A\left(\frac{\vec k}{\omega^{\a_{\vec{k}}}}\right).
\eeq
We take $\a_A=2d_U-d$.
The scaling for a Fermi liquid corresponds to
$d_U=(d-1)/2$. Because of the constraints on unparticles, $d_U$ always
exceeds $d/2-1$ for scalar unparticles and $(d-1)/2$ for unfermions, where the rule to
obtain unfermions is to set $d \to d+1$.  We term a correlated system with
such scaling an un-Fermi liquid as the basic excitations are
unparticles. Un-Fermi liquids are non-Fermi liquids composed of
unparticles, whose propagator is given by the unfermionic analogue of
Eq. (\ref{gfunparticles}),
\beq\label{unf}
S_{U}\left(\vec k\right) & \sim &
\left(\vec k^{2}-i\epsilon\right)^{d_{U}-(d+1)/2}\times\nonumber\\
&&\left(\vec {k}\!\!\!/+\cot\left(d_{U}\pi\right)\sqrt{\vec{k}^{2}-i\epsilon}\right),
\eeq
which contains a non-local mass term.
Un-Fermi liquids should not be construed as Fermi
liquids with poles at the unparticle energies.  Because the unparticle
fields cannot be written in terms of canonical ones, there is no sense
in which a Gaussian theory can be written down from which pole-like
excitations can be deduced.  Note that the scaling form, if it were
to satisfy any kind of sum rule, can only be a valid approximation
over a finite energy range.  

As pointed out by Georgi\cite{georgi}, the phase space
for $N$ massless particles is identical to the unparticle propagator but
with $d_U$ replaced by $N$ in four dimensions. In general the relationship is $d_U=N\left(d/2-1\right)$.  Since $d_U$ is in general not an integer,
interpreted loosely as an anomalous dimension, the dynamics of unparticles with
scaling dimension $d_U$ are equivalent to those of a fractional number of
massless particles. Indeed, Luttinger-liquid Green functions have been
proposed previously\cite{andersonchak,balatsky} as the source of non-Fermi liquid behaviour in (2+1)-dimensional 
systems but without justification because such Green functions have a
rigorous basis only in (1+1)-dimension.  The unparticle construction applies
regardless of $d$ and hence offers a way around this conundrum.  In addition,  when $d_U>d/2$ ($d_U>(d+1)/2$ for
fermions), the 
propagator can vanish, giving rise to zeros and an explicit violation
of the
Luttinger count. One of our key results is that $d_U$ has a natural
lower bound that guarantees that $G_U(0)=0$, leading to a realistic
model of zeros,  exploited recently in the context of overlap fermions\cite{zubkov}. 
 Consequently, we
propose here that unparticles are in part responsible for the dynamics at non-trivial IR
fixed points in strongly correlated electron matter. While in general, the
breakdown of the particle concept is well accepted to obtain at
 critical points\cite{wilson, sachdev}, our specific
proposal is that the propagator can be deduced immediately from scale
invariance\cite{georgi} and that such scaling is valid in an entire
phase (namely the pseudogap) not just at a single point, as long as zeros of the single-particle propagator persist.

\subsection{Action on AdS}

We show in this section that there is an intimate link between unparticles and the gauge-gravity duality\cite{Maldacena:1997re}. 
 In fact, the latter helps fix $d_U$ such that the unparticle propagator will always have zeros.  
We start with the action 
\beq
S_{\phi}=\frac12\int d^{4}\vec{x} \, \left(\partial_{\mu} \phi \partial^{\mu} \phi + m^2
  \phi^2\right)
\eeq
for a massive free theory, where $\phi$ is a scalar field and $m$ the
mass, and make it scale-invariant\cite{cm1,deshe} by simply integrating over mass
\beq\label{Lunparticles}
S=\int_0^\infty dm^2 B(m^2)\, S_{\phi},
\eeq
where $\phi\equiv\phi(x,m^2)$ now depends explicitly on the mass at each scale.  Here, we have included a mass-distribution function of the form $B(m^2) =a_\delta
(m^2)^\delta$%and the corresponding propagator is
%identical to Eq. (\ref{gfunparticles})
. Even though we are explicitly considering the case of scalar field here, our
conclusions apply equally to formulation with Dirac fields.

While it is common\footnote{For example, this has been used explicitly in the derivation of the unparticle propagator by
Deshpande and He\cite{deshe}.} to relate
the emergent unparticle field directly to the scalar field $\phi$
using a different mass-distribution function $f(m^2)$ through a relationship of the form
\beq\label{uphiphi}
\phi_U(\vec{x})=\int_0^\infty dm^2 \phi(\vec{x},m^2)f(m^2),
\eeq
this is not correct
as it would imply that the unparticle field $\phi_U$ has a particle
interpretation in terms of the scalar field $\phi$.  For
example, $\phi_U$ would then obey a canonical commutator and the resultant unparticle propagator could be interpreted as that of a Gaussian theory.
We demonstrate this explicitly in the Appendix.  In actuality, the unparticle field should not be a sum of $\phi(x,m^2)$s, which are independent functions of
mass, but rather should involve some unknown product of the particle fields\cite{georgi}.   Consequently, unparticle physics cannot be accounted for in a Gaussian theory.

Although $\phi_U$ and $\phi$
are not related in any straightforward way, we will show that the
action, Eq. (\ref{Lunparticles}), gives rise to a
generating functional for the unparticle strictly in the gauge/gravity duality
sense.  Indeed, some link has been noted previously between
the unparticle idea and an action on anti de Sitter (AdS) space. However,
such a connection\cite{Stephanov:2007uq,Cacciapaglia:2008kx} remains heuristic as there has been no explicit
mapping between the mass integration in the unparticle action and
the AdS metric.  It is this missing link that we provide here. The key idea is that we transform the mass into a length scale through
$m=z^{-1}$ and $z$ will appear as the radial direction in AdS.   We then deduce the generating functional for the unparticle simply by constructing the AdS on-shell action.  This will effectively remove the degree of freedom that determines the scaling dimension of the
unparticle field in the underlying AdS formulation, namely the mass of the scalar field living in the AdS space\cite{Stephanov:2007uq,Cacciapaglia:2008kx}. 

The substitution
of $m=z^{-1}$ introduces an extra factor of $z^{5+2\delta}$ into the
action as can be
seen from
\beq
\mathcal{L}&=&a_\delta\int_0^\infty
dz\frac{2R^2}{z^{5+2\delta}}\left[\frac{1}{2}\frac{z^2}{R^2}\eta^{\mu\nu}(\del_\mu
  \phi)( \del_\nu \phi)+\frac{\phi^2}{2R^2}\right]. \nonumber\\
%&=&a_\delta\int_0^\infty
%dz\frac{2L^2}{z^{d+1+\delta}}\left[\frac{1}{2}\del_a\phi\del^a\phi+\frac{\phi^2}{2R^2}\right],
\eeq
%where $a$ is a summation over the
%Minkowski coordinates.  
We propose that the correct starting point for the unparticle construction
is  thus the action on AdS$_{5+2\delta}$
\beq
S = \frac{1}{2} \int d^{4+2\delta}\vec{x} \, dz \, \sqrt{-g} \left(\partial_a \Phi \partial^a \Phi + \frac{\Phi^2}{R^2} \right).
\eeq
where 
\beq
ds^2=\frac{R^2}{z^2}\left(\eta_{\mu\nu}dx^\mu dx^\nu+dz^2\right),
\eeq
is the corresponding metric, $R$ the AdS radius,  $\sqrt{-g}=(R/z)^{5+2\delta}$, and
$\phi\rightarrow (a_\delta 2)^{-1/2}R^{3/2}\Phi$.  All of the factors of  $z^{5+2\delta}$ and the $z^2$ in the gradient
terms appear naturally with this metric.  According to the gauge-gravity duality, the on-shell action then becomes the
generating functional for
unparticle stuff which lives in an effective dimension of $d= 4+2\delta$. Therefore, we would like to have $\delta \leq 0$. Furthermore, even though it may seem that unwanted dynamics in the $z$-direction has been introduced, the solution to the equation of motion prescribed by gauge/gravity 
duality can be thought of as compensating for this, \textit{i.e.}, the solution we are substituting into the action is the one that is non-normalizable at the boundary.

That the 
unparticle propagator falls out of this construction can be seen as
follows. The equation of motion is given by
\beq
z^{d+1} \, \partial_z \left(\frac{\partial_z \Phi}{z^{d-1}}\right) + z^2 \, \partial_{\mu} \partial^{\mu} \Phi - \Phi = 0.
\eeq
For spacelike momenta $k^2 > 0$, the solutions are identical to those of Euclidean AdS. The solution that is smooth in the interior is given by
\beq
\Phi(z,\vec{x}) = \int \frac{d^{d}\vec{k}}{(2\pi)^{d}} e^{i \vec{k}\cdot \vec{x}} \frac{z^{\tfrac{d}{2}} K_{\nu} (kz)}{\epsilon^{\tfrac{d}{2}} K_{\nu} (k\epsilon)} \, \tilde{\Phi}(\vec{k}),
\eeq
where $\vec{k}$ is a $d$-momentum transverse to the radial $z$-direction and 
\beq
\nu = \frac{\sqrt{d^2 + 4}}{2}. 
\eeq
We note that this solution decays exponentially in the interior and thus, even though it is a $z$-dependent solution, one can think of $\Phi$ as localized at the boundary $z=\epsilon \rightarrow 0$.  Here, we have explicitly cut off the AdS geometry to regularize the on-shell action
\beq
S &=& \frac{1}{2} \int d^{d}\vec{x} \, g^{zz} \sqrt{-g} \,\Phi(z,\vec{x}) \partial_z \Phi(z,\vec{x}) \Big|_{z= \epsilon} \nonumber \\
&=& \frac{1}{2} \frac{R^{d-1}}{\epsilon^{d-1}} \int
\frac{d^{d}\vec{p}}{(2\pi)^{d}}\frac{d^{d}\vec{q}}{(2\pi)^{d}}
\, (2\pi)^{d} \delta^{(d)}(\vec{p}+\vec{q})\nonumber\\
&& \tilde{\Phi}(\vec{p}) \frac{d}{d\epsilon} \left(\log \left[\epsilon^{\tfrac{d}{2}} K_{\nu}(p\epsilon)\right]\right) \tilde{\Phi}(\vec{q}). \nonumber \\
\eeq
Interpreting this as a generating functional for the unparticle field
$\Phi_U$ living in a $d$-dimensional spacetime, we
can then read the (regulated) 2-point function, which scales like
$p^{2\nu}$. We can then  analytically continue to the case of timelike momenta, which corresponds
 to choosing the non-normalizable solution to the bulk equation of motion. This analytically continued solution will also be localized at the boundary. 

The 2-point function of the unparticle in real space is then given by
\beq
\langle \Phi_U (\vec{x}) \Phi_U(\vec{x}') \rangle = \frac{1}{|\vec{x} -\vec{x}'|^{2d_U}},
\eeq
where 
\beq\label{du}
d_U = \frac{d}{2} + \frac{\sqrt{d^2 + 4}}{2} > \frac{d}{2},
\eeq
and $d$ is the dimension of the spacetime the unparticle lives
in.  We note that in this construction, there is only one possible scaling dimension for the unparticle instead of two, due to the fact that the square of the mass of the AdS scalar field $\Phi$ is positive\footnote{In AdS, a stable particle is allowed to have negative mass squared as long as it satisfies the so-called Breitenlohner-Freedman bound.}. As a result, the
unparticle propagator has zeros defined by
$G_U(0)=0$, not infinities. This is the principal result of this construction. 

The AdS interpretation of unparticle we propose here is different
than those in Refs. \cite{Stephanov:2007uq,Cacciapaglia:2008kx} and the
differences are as follows.  In Refs. \cite{Stephanov:2007uq,Cacciapaglia:2008kx}, the spacetime dimension of the AdS space is $5$, while here, the spacetime dimension is $d+1 = 5 + 2\delta$. 
Second, the mass $m_{AdS}$ of the particle living in the AdS interpretation of Refs. \cite{Stephanov:2007uq,Cacciapaglia:2008kx} is a free parameter which is related to the scaling dimension of the unparticle field. In particular, 
\beq
m_{AdS}^2 = \frac{d_U (d_U - 4)}{R^2}.
\eeq
In our interpretation, however, $m_{AdS}^2 = 1/R^2$. Third, since $m_{AdS}$
is a free parameter, the scaling behavior of the 2-point function of
the unparticle field is undetermined in
Refs. \cite{Stephanov:2007uq,Cacciapaglia:2008kx}, while in our case,
it depends solely on the dimension  of the spacetime the unparticle
lives in. 

%In our AdS construction,  we have $d = 4$ regardless of $d_{\phi}$, and this actually fixes the
%unparticle scaling dimension to 
%\beq
%d_U = 2+\sqrt{5} \approx 4.24.
%\eeq  
%We note that in other unparticle interpretations, such as those in Refs. \cite{georgi,jimmy}, there is no restriction on the dimensionality of the spacetime the unparticle lives in. For the remainder of this paper, we will revert back to the set-up with generic spacetime dimensions.

\subsection{Statistics}
  
Our proposal that unparticles
are the fundamental excitations in strongly correlated systems (at least ones that possess zeros) has a key experimental prediction.
As pointed out by Georgi\cite{georgi}, unparticles have a propagator
equivalent to that of  $N$ massless particles where $N=d_U/(d/2-1)$.
Consider the case of $d=2+1$.  Extending the arguments used in the
context of anyons\cite{klresponse,wilczek,halperin}, we find that because of the branch-cut structure of
the propagator, there should be a non-trivial phase upon unparticle
exchange given by $e^{i2\pi d_U}\ne \pm 1$ that is directly related to the
fact that interchange of unparticles amounts to an interchange of
$d_U/(d/2-1)$ massless particles. Since $d_U$ is non-integer (see Eq. (\ref{du})), any statistics
are possible.  Consequently, clockwise and counterclockwise rotations
of unparticles do not yield the same phase, thereby indicating a spontaneous
 time-reversal
symmetry breaking (TRSB). The TRSB found here for
unparticles arises fundamentally from the interactions that lead to the non-trivial excitations
in the IR and hence avoids the argument against TRSB based on
quasiparticles\cite{kivelson}.  Our arguments apply strictly to $d=2+1$ and hence if
applicable to doped Mott insulators are relevant only to a single
copper-oxide layer. Certainly subtleties will arise in applying them
to bulk 3-dimensional materials.  Nonetheless, it is interesting to note that numerous
experiments\cite{trsb1,trsb2,trsb3} have reported observations
consistent with the breaking of time-reversal symmetry in the
pseudogap phase (a phenomenon distinct from the surface-induced
breaking of time reversal symmetry in the superconducting
state\cite{trsb11}).  However, recent work\cite{ornstein,kivel} has
suggested that Kerr effect\cite{trsb3} measurements are more
consistent with the
breaking of inversion symmetry rather than time-reversal because the
signal fails to change sign when a measurement is made on the opposite
surface. Certainly all of the anyon
constructions\cite{wilczek,halperin,klresponse} require both
inversion and time-reversal symmetries to be broken.  However, since scale-invariant matter does not require inversion
symmetry breaking, it seems unlikely that the non-trivial statistics
associated with unparticles in $d=2+1$ 
would result in a sign change of the Kerr signal for measurements on
opposing surfaces. Consequently, there exists no ${\it a priori}$ contradiction between the unparticle construction and the Kerr effect\cite{trsb3} observations.  Hence, experiments designed to search for
non-trivial statistics in the pseudogap regime are most relevant here.
Since unparticles yield zeros, they can be localized and hence could
be interchanged thereby making a direct measurement of their
statistics possible.  Experiments along these lines would certainly be sufficient to
falsify the relevance of unparticles to pseudogap matter.  Nonetheless, the zero feature
of the unparticle propagator is noteworthy because it can explain the
dip in the density of states (that is the pseudogap) a feature which
is absent in other work\cite{trsb4} which can explain just the presence of potential TRSB.   Regardless of the applicability
of these results to the experiments\cite{trsb1,trsb2,trsb3}, this
works indicates that TRSB can quite generally arise from the strong
correlations that remain from the Mott state, that is, Mottness, thereby offering a realization of the general principle underlying interaction-induced fractional statistics advocated by Jones-Marshall and Wilczek\cite{wilczek}.

\section{Superconducting Instability}

Because unparticles do not have any particular energy, they should be
useful in describing physics in which no coherent quasiparticles
appear, as in the normal state of the cuprates.  Since all
formulations of superconductivity start with well-defined
quasiparticles, we explore what happens when we use a quasiparticle spectral function with a scaling form. 
Such an approach is warranted given that the cuprates
 exhibit a color change\cite{marel1,rubhaussen} upon a transition to
 the superconducting state as
 evidenced most strikingly by the violation of the
 Ferrell-Glover-Tinkham sum rule\cite{marel1}. Some initial work along these lines has been proposed
previously\cite{sczhang,balatsky,andersonchak}, which have all been
based on the Luttinger-liquid Green function. As remarked earlier, the
advantage of the unparticle approach is that it is completely general
regardless of the spatial dimension unlike the Luttinger-liquid
one which must be restricted to $d=1+1$.  To obtain the general
result, we work first with the scaling form of the spectral function as in
Eq. (\ref{specfun}), generalizing the procedure in \cite{balatsky}.  

Since the machinery to deal with pairing instabilities with non-trivial statistics does not exist for arbitrary anomalous dimension, $d_U$, our goal in this section is to see if something new arises in the superconducting
instability when fermionic particles are described by a scale-invariant spectral function of the form of Eq. (\ref{specfun}).   We take a
system of fermions that have a separable two-body interaction $V\left(\vec{k}-\vec{k}^{\prime}\right)=\lambda w_{\vec{k}}^{*}w_{\vec{k}^{\prime}}$, but
 are described by a spectral function that has an effective scaling form up to some
energy scale $W$ and some lower bound close to zero. The equation for the existence of an instability
in terms of the Green function is

\begin{eqnarray}
1 & = & i\lambda\sum_{\vec{k}}\left|w_{\vec{k}}\right|^{2}G\left(\vec{k}+\vec{q}\right)G\left(-\vec{k}\right).
\end{eqnarray}
This gives the zero temperature result. We take the interaction strength
$\lambda$ as a constant of mass dimension $2-d$, and $w_{\vec{k}}$ as
a filling factor. We work in the center of mass frame, such that $\vec{q}=\left(q_{0},0\right)$.
The critical temperature for a second-order transition corresponds
to $q_{0}=0$. Then switch to imaginary time to work at finite temperature,
so that the new equation reads
\begin{eqnarray}
\label{crittempconstraint}
1 & = & \lambda T\sum_{n,\vec{k}}\left|w_{\vec{k}}\right|^{2}G\left(\omega_{n},\vec{k}\right)G\left(-\omega_{n},-\vec{k}\right).
\end{eqnarray}
The Green function is related to the spectral function via

\begin{eqnarray}
G\left(\omega_{n},\vec{k}\right) & = & \int_{-\infty}^{\infty}dx\frac{A\left(x,\vec{k}\right)}{x-i\omega_{n}}.
\end{eqnarray}
Then we obtain, using $\omega_{n}=\pi T\left(2n+1\right)$,

\begin{eqnarray}
1 & = & \frac{\lambda}{2}\int dxdy\sum_{\vec{k}}\left|w_{\vec{k}}\right|^{2}A\left(x,\vec{k}\right)A\left(y,-\vec{k}\right)\nonumber\\
&&\times\frac{\tanh\left(x/2T\right)+\tanh\left(y/2T\right)}{x+y}.
\end{eqnarray}
The $\vec{k}$-dependence is recast in terms of $\xi\left(\vec{k}\right)$, which
is in general some function of $\vec{k}$ with units of energy, which for
instance can always be done for an isotropic system. In BCS, $\xi\left(\vec{k}\right)$
would correspond to kinetic energy. Take $\sum_{\vec{k}}\left|w_{\vec{k}}\right|^{2}\to \left(\text{Volume}\right)^{-1}\times N\left(0\right)\int d\xi$
so we pull out a constant density of states. The integral is now rewritten
as
\begin{eqnarray}
1 & = & \frac{g}{2}\int dxdy\int_{0}^{\omega_{c}}d\xi A\left(x,\xi\right)A\left(y,\xi\right)\nonumber\\
&&\times\frac{\tanh\left(x/2T\right)+\tanh\left(y/2T\right)}{x+y}
\end{eqnarray}
where $\lambda N\left(0\right)\times\left(\text{Volume}\right)^{-1}=g$ such that $g$
is a dimensionless measure of interaction strength. Evaluating the integral at any temperature gives the minimal coupling
to cause a pairing instability, and this equation traces out a phase diagram for $g$ and $T$. 

We can extract some qualitative features knowing the spectral function
is attenuated at low and high energy. In the limit of extremely low
critical temperature, the entire range of nonzero $A$ falls in the
region where $\tanh\left(x/2T\right)\approx1$, and therefore the
integral becomes independent of critical temperature to this order,
so $g\approx\text{const}$ (in the case of BCS, this would just be
$0$ as $T\to0$ due to the logarithmic divergence from  the integral).
Oppositely in the limit of very high critical temperature, over the
entire region it is valid to take $\tanh\left(x/2T\right)\approx x/2T$,
thus $g\approx4T/\omega_{c}$. Thus regardless of the specific form
of the spectral function, $g$ will increase with critical temperature
for high temperatures. 

Now we look in the intermittent region of energy where we take advantage
of the scaling form. The ``beta function'' looks like

\begin{eqnarray}
\label{constraintTderivative}
\frac{dg}{d\ln T} & = & \frac{g^{2}}{4T}\int dxdy\int_{0}^{\omega_{c}}d\xi A\left(x,\xi\right)A\left(y,\xi\right)\nonumber\\
&&\times\frac{x\, \text{sech}^{2}\left(x/2T\right)+y\, \text{sech}^{2}\left(y/2T\right)}{x+y}.
\end{eqnarray}
We can recover the BCS result using $A\left(\omega,\xi\right)=\delta\left(\omega-\xi\right)$,

\begin{eqnarray}
\frac{dg}{d\ln T}  =  g^{2}\tanh\frac{\omega_{c}}{2T} \approx g^2.
\end{eqnarray}

The $\text{sech}^{2}$ terms in Eq. (\ref{constraintTderivative}) exponentially suppress high-energy contributions.
With an appropriate attenuation on the low-energy side from the spectral function, it is possible for the beta function to be negative seeing how $\left(x\text{sech}^{2}x+y\text{sech}^{2}y\right)/(x+y)$ dips into negative values, which are most pronounced along the line $x+y=0$. In addition these values can outweigh ones for the corresponding positive entries along $x-y=0$. For instance, the smallest minimum of the
function is at $(x,y)=\left(1.35,-1.35\right)$ and we obtain a value
of $-0.321$ but for $(x,y)=\left(1.35,1.35\right)$ we find $0.235$.
With appropriate weight and suppression, the negative values can dominate
the integral and confer a total negative sign. This suppression seems natural in the realm of $\a_A>0$, where the scaling form naturally takes on smaller values at lower energies. This would mean that
coupling strength increases with decreasing temperature, and therefore
there is a bottoming out.  That is, a minimum coupling strength that can confer
superconductivity exists. 

Let us more closely examine what happens when we impose the scaling
form at the outset.  Then approximately

\begin{eqnarray}
1 & = & \frac{g}{2}\tilde{T}^{2\left(1+\alpha_A\right)}\int dxdy\int_{0}^{\omega_{c}/\tilde{T}}d\xi A\left(x,\xi\right)A\left(y,\xi\right)\nonumber\\
&&\times\frac{\tanh\left(x/2W\right)+\tanh\left(y/2W\right)}{x+y}
\end{eqnarray}
where the tilde denotes the ratio of that energy to $W$, \emph{e.g.} $\tilde{T}\equiv\frac{T}{W}$.
The scaling form of the spectral function confers a scaling form for $g$ like

\begin{eqnarray}
g\left(\tilde{T},\tilde{\omega}_{c}\right) & = & \tilde{T}^{-2\left(1+\alpha_A\right)}f_g\left(\frac{\tilde{T}}{\tilde{\omega}_{c}}\right).
\end{eqnarray}
Now sequentially we take a logarithm, derivative and finally rescale
the remaining integral back to obtain

\begin{eqnarray}
\frac{dg}{d\ln\tilde{T}} & = & -2\left(1+\alpha_A\right)g+\frac{g^{2}}{2}\omega_{c}\int dxdyA\left(x,\omega_{c}\right)A\left(y,\omega_{c}\right)\nonumber\\
&&\times\frac{\tanh\left(x/2T\right)+\tanh\left(y/2T\right)}{x+y}.
\end{eqnarray}
The second term is positive-definite. This term can conceivably be
small if there is relatively little spectral weight near $\omega_{c}$
within the scaling form or, equivalently, that $g$ is not very susceptible to changes in $\omega_c$. In this event, then we have 
\beq
\frac{dg}{d\ln\tilde{T}} =  -2\left(1+\alpha_A\right)g+O\left(g^{2}\right).
\label{betafcn}
\eeq
The right-hand side of this expression is strictly negative for our region of interest where $\alpha_A>0$.  Hence, we find quite generally that the critical temperature increases as
the coupling constant decreases! 

\begin{figure}[h!]
\begin{center}
\includegraphics[width=3.0in]{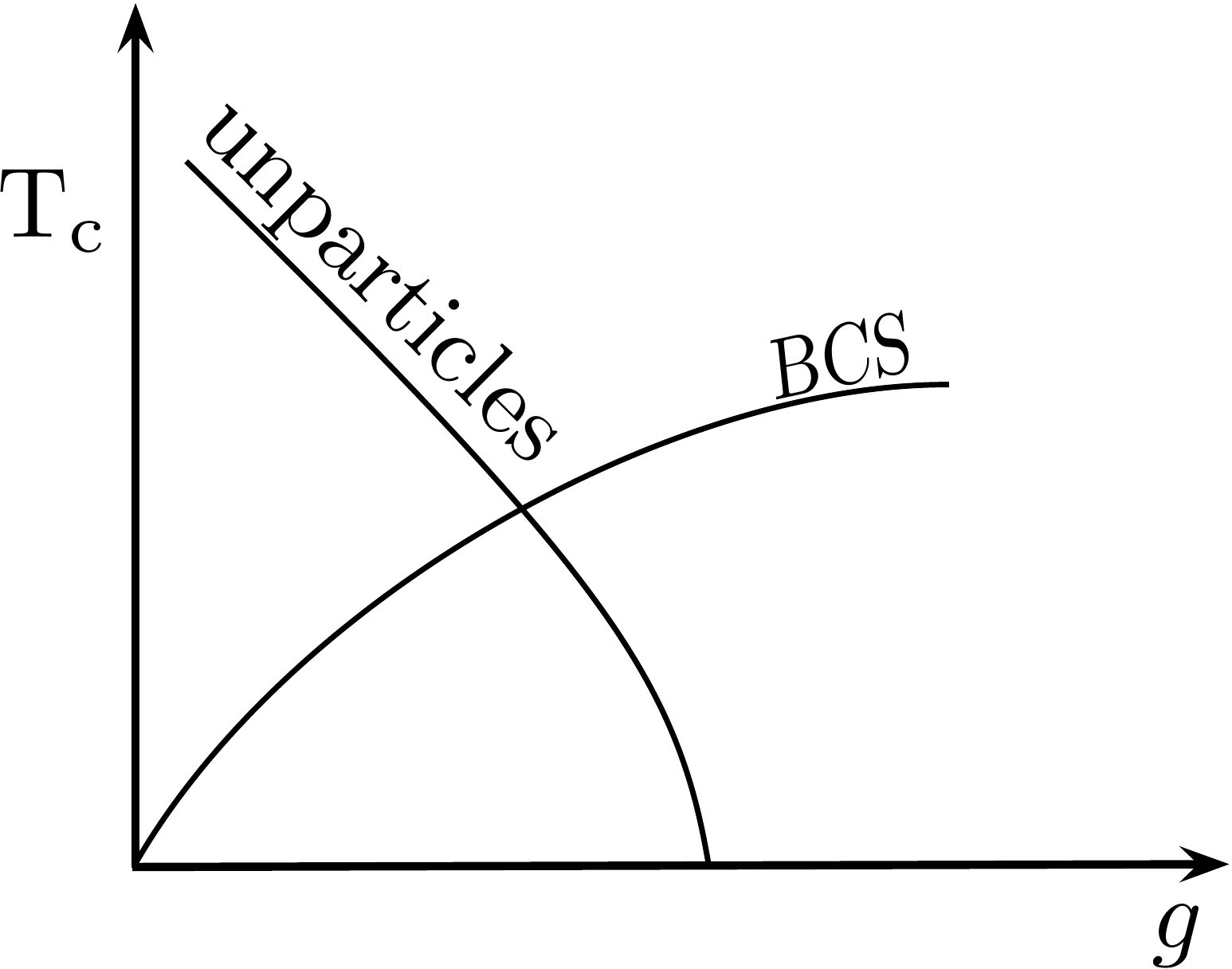}
\caption{  Plot of the $\beta$-function for the superconducting
  transition in the ladder approximation for unfermions,
  Eq. (\ref{betafcn}).  We have considered the case $d_U>d/4$ for the scaling dimension
  of the unparticle field, a condition naturally satisfied in the AdS construction of unparticles.  The contrast with the standard prediction
  of BCS theory is shown. }
\label{TC}
\end{center}
\end{figure}

This stands in stark contrast to the Fermi
liquid case in which just the opposite state of affairs obtains.  
This is illustrated clearly in Fig. (\ref{TC}).  In the context of the
cuprate superconductor problem, the opposing trends for $T_c$ versus
the pairing interaction suggests that perhaps a two-fluid model
underlies the shape of the superconducting dome assuming, of course, that a similar behaviour for $T_c$ as a function of doping persists\footnote{There is of course no connection between $g$ and $x$. Certainly a generalization of this work to a doped model that admits unparticles could be studied as a function of doping to see if the qualitative trends in Fig. (\ref{TC}) obtain.}Since the
transition to the superconducting state breaks scale invariance, the
particle picture should be reinstated.  Consequently, we expect the
broad spectral features dictated by the branch cut of the unfermion
propagator to vanish and sharp quasiparticle features to appear upon
the transition to the superconducting state as is seen experimentally\cite{he}.  

\section{Closing}

We have proposed that using scale invariance as an organizing
principle aids in computing the properties of strongly correlated electron systems at low energy.  The necessity for a
non-Gaussian fixed point arises anytime the single-particle Green
function vanishes. Such a vanishing obtains at either a single
point or in an entire phase, such as in the pseudogap phase of the
cuprates.  Our key proposal here is that scale invariance persists
as long as the single-particle Green function vanishes over a locus of points in momentum space.   The key signature
of this behavior which is testable experimentally is the critical
scaling of the spectral function in Eq. (\ref{specfun}) in the entire region. 

We have also proposed an interpretation of unparticles using the AdS
construction, which permits us to fix the scaling dimension of the unparticle field and ensures that its propagator will have zeros.  Since this propagator
possesses zeros, it is a candidate to explain the breakdown of Fermi
liquid theory from strong interactions. 

There is a simple way of understanding why a divergent self-energy results in excitations which have fractional statistics.
A divergent self energy represents an orthogonality catastrophe, implying that the underlying excitations have no overlap with
the starting particle fields.  Hence, some new fundamental objects which have no canonical particle interpretation carry the charge.  The excitations that emerge are composites.  In $d=2+1$, the new excitations
 can acquire
non-trivial exchange statistics.  While experiments to detect fractional statistics are
notoriously difficult, we hope this work provides the impetus to
search for them in the pseudogap phase of the cuprates, where the zeros of the
propagator exist.  Note the zeros give rise to localised excitations and hence unparticle interchange is certainly feasible experimentally.  Two other predictions that are
falsifiable experimentally are 1) a spectral function exhibiting the
scaling of Eq. (\ref{specfun}) and 2) a deviation from the  Luttinger
count, the latter having already received experimental confirmation\cite{dave}.  The latter prediction puts this theory in direct contrast with the leading phenomenological theory
of the pseudogap regime in which the Luttinger count is strictly maintained\cite{yrz}.

Finally, we have shown that in contrast to the Fermi liquid case, the
branch cut singularity in the unparticle propagator gives rise to a superconducting instability in which the critical temperature increase as the coupling constant decreases. This suggests that perhaps a two-fluid model underlies the shape of the superconducting dome of the cuprate superconductors.

\textbf{Acknowledgements}
P. Phillips thanks M. Zubkov for the e-mail communication
 pointing out the connection between unparticles and zeros, S. Balatsky for a clarifying exchange on Ref. 26, T. Hughes for comments on the statistics section of the paper which led to our emphasis on time-reversal symmetry breaking and N. P. Armitage for an exchange on time-reversal symmetry breaking.  The work of B. Langley and P. Phillips is supported by NSF DMR-1104909 which grew out of earlier work funded by the
Center for Emergent Superconductivity, a DOE Energy Frontier Research Center, Grant No.~DE-AC0298CH1088. J. Hutasoit is supported by NSF grant DMR-1005536 and DMR-0820404 (Penn State MRSEC).

\section{Appendix}
\begin{widetext}
We show here that if only if a linear relationship between the unparticle and particle fields is maintained (as in Eq. (\ref{uphiphi})), then a Gaussian action for the unparticles obtains.
Let us turn the action in terms of the massive fields into an action
in terms of unparticle fields. The original partition function is
given by
\begin{eqnarray*}
\mathcal{Z} & = & \int\mathcal{D}\phi_{n}e^{i\int d^{d}p\mathcal{L}\left[\left\{ \phi_{n}\right\} \right]}\\
\mathcal{L} & = & \frac{1}{2}\sum_{n}B_{n}\phi_{n}\left(p\right)\left(p^{2}-M_{n}^{2}\right)\phi_{n}\left(-p\right).
\end{eqnarray*}
where $n$ is to indicate a sum over the mass $M_{n}$. This sum can remain a general sum over various free fields, but we will ultimately take the limit where the sum is a continuous sum over all masses. The factor $B_n$ is a weight factor that, in the continuous mass limit, will change the mass dimension of $\phi_n$. We introduce
a Lagrange multiplier through a factor of unity and simply integrate
over the all fields that are not $\phi_{U}$ to obtain

\begin{eqnarray*}
\mathcal{Z} & = & \int\mathcal{D}\phi_{n}\mathcal{D}\phi_{U}\mathcal{D}\lambda\exp\left\{ i\int d^{d}p\left(\frac{1}{2}\sum_{n}B_{n}\left(p^{2}-M_{n}^{2}\right)\phi_{n}^{2}+\lambda\left(\phi_{U}-\sum_{n}F_{n}\phi_{n}\right)\right)\right\} \\
 & = & \int\mathcal{D}\phi_{U}\mathcal{D}\lambda\exp\left\{ i\int d^{d}p\left(\lambda\phi_{U}-\frac{1}{2}\lambda^{2}\sum_{n}\frac{F_{n}^{2}}{B_{n}\left(p^{2}-M_{n}^{2}\right)}\right)\right\} \\
 & = & \int\mathcal{D}\phi_{U}\exp\left\{ \frac{i}{2}\int d^{d}p\phi_{U}\left(p\right)\left(\sum_{n}\frac{F_{n}^{2}}{B_{n}\left(p^{2}-M_{n}^{2}\right)}\right)^{-1}\phi_{U}\left(-p\right)\right\} 
\end{eqnarray*}
with repeated absorptions of normalization contants into the measure. The factor $F_n$ is another weight factor, this time chosen to determine the scaling dimension of the unparticle field $\phi_U$. Bescause $F_n$ is chosen to give $\phi_U(x)$ a scaling dimension $d_U$, in the continuous mass limit the ratio $F_n^2/B_n \sim \left(M_n^2\right)^{d_U-\frac{d}{2}}$. This is necessary because of how $F_n$ imposes the scaling dimension.
Hence we identify the propagator of the unparticle field as
\begin{eqnarray*}
G_{U}\left(p\right) & = & \sum_{n}\frac{F_{n}^{2}}{B_{n}\left(p^{2}-M_{n}^{2}\right)}\\
 & \sim & \left(p^{2}\right)^{d_{U}-\frac{d}{2}}.
\end{eqnarray*}

This argument can also be run in reverse.  Namely, if we assume a Gaussian action for the unparticles then the Lagrange multiplier constraint in the form of Eq. (\ref{uphiphi}) is implied.  
\end{widetext}

\end{document}